\documentclass[pdflatex,sn-mathphys]{sn-jnl}

\usepackage{amsmath}
\usepackage{amsfonts}
\usepackage{bm}
\usepackage{physics}
\usepackage[utf8]{inputenc}

\jyear{2021}%

\theoremstyle{thmstyleone}%
\theoremstyle{thmstyletwo}%

\theoremstyle{thmstylethree}%

\newcommand{\av}[1]{\left \langle #1 \right\rangle}

\begin{document}
	
	\title[Insights into Quantum Contextuality and Bell--Nonclassicality]{Insights into Quantum Contextuality and Bell Nonclassicality: a study on Random Pure Two-Qubit Systems}
	%
	
	
	
	\author[1]{\fnm{Giovanni} \sur{Scala}}\email{giovanni.scala@ug.edu.pl}
	
	\author[1]{\fnm{Antonio} \sur{Mandarino}}\email{antonio.mandarino@ug.edu.pl}
	
	\affil[1]{\orgdiv{International Centre for Theory of Quantum Technologies}, \orgname{University
			of Gdansk}, \orgaddress{\street{Jana Ba\.zy\'nskiego 1A}, \city{Gda\'nsk}, \postcode{80-309}, \country{Poland}}}

	
	\abstract{We explore the relationship between Kochen-Specker quantum contextuality and Bell-nonclassicality for ensembles of two-qubit pure states. We present a comparative analysis showing that the violation of a noncontextuality inequality on a given quantum state reverberates on the Bell-nonclassicality of the considered state. 
		In particular, we use suitable inequalities that are experimentally testable to detect quantum contextuality and nonlocality for systems in a Hilbert space of dimension $d=4$. 
		While contextuality can be assessed on different degrees of freedom of the same particle, the violation of local realism requires parties spatially separated.}
	

	
	
	
	\maketitle
	
	\section{Introduction}\label{sec1}
	The possibility of explaining the results obtained in quantum mechanics within a theory 
	admitting hidden variables has been excluded via two theorems. The Bell-Kochen-Specker theorem \cite{RBell_C, KS_th}, 
	which is the most general, and the  theorem of John Bell \cite{Bell_th}. 
	The first one discards all the noncontextual hidden variable 
	(NCHV) theories, namely all those in which physical quantities have a predefined outcome value, 
	no matter the experimental context. A context is defined by the
	other comeasurable observables that are measured simultaneously with a given one. 
	The formalism of quantum mechanics imposes that two observables, $A$ and $B$, 
	are comeasurable if their associated operators commute, i.e. $ [ A, B ]:= AB - BA = 0 $.
	The second theorem rules out local hidden variable (LHV) theories. 
	Originally, Bell employed a  locality argument in formulating his theorem. He postulated that if two systems lie outside each other's light cones, any measurement conducted on one system will not impact the other. This locality assumption is a fundamental constraint for these theories. Consequently, the outcomes of measurements on spatially separated systems remain unaffected by the specific selection of measurement settings in a different, spacelike-separated region of space-time.
	
	The conflict between the assumptions of quantum mechanics and the concept of locality can be demonstrated in two distinct ways. One way is through the violation of an inequality that local hidden variable (LHV) models would normally satisfy. The other way is by revealing a logical contradiction between the predictions of quantum mechanics and those of LHV models.
	
	Specifically, the first approach, based on inequality violation, provides remarkable and empirically testable cases. For instance, the Clauser-Horne-Shimony-Holt (CHSH) inequality \cite{CHSH} is defined in terms of two-point correlation functions. On the other hand, the more comprehensive Clauser-Horne (CH) inequality \cite{CH74} is expressed in terms of joint probabilities. For scenarios involving more than two parties, other examples of inequalities can be found in \cite{Zukowski2002,Karczewski2022, Bae2018}.

	The main difference between the Bell-Kochen-Specker theorem and the Bell's one 
	is the number of systems they involve. In fact, the proof of the former 
	refers to a single individual system but requires carrying out measurements of 
	non-compatible observables that cannot be simultaneously measured on an individual system. 
	Whereas the proof of the Bell theorem, in the form of inequality, 
	involves the outcome statistics of measurement performed on different separated subsystems. 
	Moreover, Bell inequalities are satisfied by any LHV theory and 
	they are independent of the assumptions needed in quantum mechanics \cite{cabello08}. 
	
	In contrast, the original proofs of the Bell-Kochen-Specker theorem 
	refer to NCHV theories that share some properties with the formulation of quantum theory. 
	Consequently, this enables two distinctive tests of the contextual nature of quantum mechanics, 
	generally referred to as state-independent, as in the argument of Peres-Mermin square, 
	and state-dependent contextuality. 
	The primary distinction between these tests lies in their sensitivity to state preparation and the considered measurement scenario. The state-independent test reveals a contradiction between noncontextual hidden variable (NCHV) theories and quantum mechanics, regardless of the state's preparation. In contrast, the state-dependent test is scenario-sensitive. The most well-known of these tests include the KCBS inequality for qutrits \cite{KCBS} and the CHSH \cite{CHSH}, both of which form part of a broader range of inequalities associated with compatibility structures.
	
	The aim  of this paper is to link tests of quantum contextuality to the deeper explored field of Bell-nonclassicality. 
	The latter has been proven that constitutes the essential resource for applications such as device-independent protocols \cite{DeviceInd_rev}, 
	and while understanding the relevance of the former to quantum computation and information processing is still in its infancy, see \cite{budroni2022kochen} for a thorough discussion. 
	In fact, possible new forms of quantum information technologies involving the least number of physical systems, 
	such as cryptographic protocols \cite{Ekert1991,hyperbit} and sensing \cite{Pepe_2022}, 
	require a deeper understanding of the quantum resources fueling their advantage over classical approaches \cite{1stPaper, 1stPLA}. 
	This research aiming at characterizing robust and trustful certification  of quantumness with a single particle is an active field of research \cite{2ndPaper, CommentDun, Catani2022a, Catani2022b}.

	\section{Kochen-Specker--like noncontextuality inequalities}\label{sec:ineq}
	In this section, we review the basic concept of quantum contextuality in the sense of Kochen-Specker and Bell-nonclassicality, 
	focusing our attention on tests involving systems belonging to the Hilbert space $\mathbb{C}^4\simeq \mathbb{C}^2\otimes \mathbb{C}^2$. 
	
	\subsection{Tests of quantum contextuality \emph{\`a la} Cabello}
	The simplest argument that manifests Kochen-Specker contextuality has been introduced by 
	Peres and Mermin \cite{peres1990incompatible, mermin90, peres1991two, peres1992recursive}. This approach is known as the Peres-Mermin (PM) square, due to the fact that 
	the nine considered measurements are arranged in a $3 \times 3 $ square. All measurements are dichotomic 
	(i.e., they have only two possible results $-1$ or $+1$) and 
	a \emph{context} is given by the product of three measurements in a row or in a column. The PM square then reads:
	\begin{equation} \mathrm{PM} =
		\begin{bmatrix}
			O_{11}       & O_{12} & O_{13}  \\
			O_{21}       & O_{22} & O_{23} \\
			O_{31}       & O_{32} & O_{33} \\
		\end{bmatrix}
	\end{equation}
	Any classical NCHV theory assigns a preexisting value to the outcomes independently of the state preparation and context. 
	Therefore we can introduce the first inequality based on the $\mathrm{PM}$ square: 
	\begin{eqnarray}
		\label{pmsquare}
		\mathrm{PM}  &=& \langle O_{11} O_{12} O_{13} \rangle +\langle O_{21} O_{22} O_{23} \rangle +\langle O_{31} O_{32} O_{33} \rangle \nonumber \\
		&+&    \langle O_{11} O_{21} O_{31} \rangle +\langle O_{12} O_{22} O_{32} \rangle  - \langle O_{13} O_{23} O_{33} \rangle \leq 4, 
	\end{eqnarray}
	where $$\langle O_{ij}  O_{kl}  O_{mn} \rangle \equiv \text{Prob}[O_{ij}  O_{kl}  O_{mn} = 1 ] - \text{Prob}[O_{ij}  O_{kl}  O_{mn} = - 1 ] $$ 
	is computed by preparing an ensemble of systems in the same state and measuring the three quantities.  
	It is easy to prove that in any \emph{classical} theory, the bound 4 is never violated \cite{cabello08}. 
	Moving to the quantum domain, one can associate an observable to each measurement $O_{ij}$, 
	with the condition that if two observables share a subindex, then they are compatible. 
	Thence, introducing $\sigma_\alpha$ with $\alpha=\{0, x, y, z\}$ ($\sigma_0$ denotes the identity operator) 
	in the case of a two-qubit system, one has:  
	
	\begin{equation} PM =
		\begin{bmatrix}
			\sigma_z \otimes  \sigma_0      & \sigma_0 \otimes  \sigma_z & \sigma_z \otimes  \sigma_z  \\
			\sigma_0 \otimes  \sigma_x     & \sigma_x \otimes  \sigma_0 & \sigma_x \otimes  \sigma_x \\
			\sigma_z \otimes  \sigma_x       & \sigma_x \otimes  \sigma_z & \sigma_y \otimes  \sigma_y  \\
		\end{bmatrix}
	\end{equation}
	and the expectation value on a state $ \ket{\psi}$ is given by 
	\begin{equation}
		\langle O_{ij}  O_{kl}  O_{mn} \rangle = \bra{\psi} O_{ij}  O_{kl}  O_{mn} \ket{\psi}.
	\end{equation}
	
	The computation of the expectation value of the product of observable on either a row or a column straightforwardly yields a result of 1, with the exception of the third column, which provides a result of $-1$. Consequently, this leads to an expectation value of $ PM  = 6$ for each quantum mechanical state under consideration. Notably, experimental findings consistently align with the quantum mechanics prescription \cite{Exp_context1, Exp_context2}. Additionally, this principle has been recently expanded to encompass bosonic fields \cite{Non_contexAM}.

	To summarize, the PM inequality in \eqref{pmsquare} is derived under the assumption that the value of the outcomes of the nine observables 
	are independent of the context in which they are measured, hence its violation implies that an \emph{a priori} value cannot be assigned, thereby underscoring the role of context in the measurement process.
	
	More recently, Cabello and coauthors in a series of papers \cite{cabello98, cabello08, cabello_exp08} presented several inequalities 
	that are experimental-friendly and that can detect quantum noncontextuality \cite{cabello_exp09}.  
	We revise here his reasoning briefly. 
	Let us consider again dichotomic observables as in the PM square case and with the same definition of compatibility based on a shared subindex. 
	Cabello proved that in any NCHV theory, the following inequality holds: 
	
	\begin{eqnarray}
		\label{Cabsquare}
		&-&\langle O_{12} O_{16} O_{17} O_{18} \rangle - \av{O_{12} O_{23} O_{28} O_{29}} - \av{O_{23} O_{34} O_{37} O_{39}} \nonumber \\
		&-& \av{O_{34} O_{45} O_{47} O_{48}}  \av{O_{45} O_{56} O_{58} O_{59}} - \av{O_{16} O_{56} O_{67} O_{69}} \nonumber \\
		&-& \av{O_{17} O_{37} O_{47} O_{67}} - \av{O_{18} O_{28} O_{48} O_{58}} - \av{O_{29} O_{39} O_{59} O_{69}}     \leq 7.
	\end{eqnarray}
	The maximal classical value is computed considering all the possible values of the product of the 18 observables, and then 
	if we can measure it on different systems, the average satisfies the bound in \eqref{Cabsquare}. 
	It is worth underlining that the product cannot be measured on a single system because it contains observables that are not compatible.
	Anyway, the fundamental assumption is that the result of any of the observables  $O_{ij}$ is independent of the context in which it is measured. 
	So,  one can measure during an experimental run a subset of compatible observables on equally prepared states.
	In the case of a physical system of dimension $d=4$, the quantum mechanical prediction of the left side of \eqref{Cabsquare} 
	is 9, independently of the state, and that is indisputably violating the value obtained within a noncontextual theory. 
	
	\subsection{CHSH-Bell inequality}
	A similar derivation of the inequality in Eq.\eqref{Cabsquare}, but assuming that the result of a measurement of an observable 
	$O_{ij}$ does not dependent on spacelike separated measurements and not on compatible measurements, 
	leads to the well celebrated CHSH-Bell inequality \cite{CHSH, gigena2022a, entropy_FID_CHSH}.
	In this case, the measurements are also dichotomic, but one has to identify spatially separated observers acting on a different subsystem.
	We denote the observables of the first part by $A_i$ and those of the second by $B_i$, and
	we can introduce the \emph{Bell--CHSH functional} as:
	\begin{equation}
		\label{eq:Bval}
		\mathcal{B}= \av{A_1 B_1} +\av{A_1 B_2} +\av{A_2 B_1} - \av{A_2 B_2}. 
	\end{equation}
	This expression constitutes a necessary and sufficient condition to establish if a state admits a description in terms of 
	a LHV model derived under the assumption of local realism. In fact, all they are constrained by $\mathcal{B} \leq 2.$ 
	A similar bound can be derived also for all quantum states, there are no quantum states that can have a value $\mathcal{B} \geq 2\sqrt{2}$, 
	this value is often referred to as Tsirelson bound \cite{Cirelson1980}.

	\section{Quantum contextuality vs Bell-Nonclassicality}\label{sec:results}
	In this section, we compare the violations of two inequalities, having different aims. 
	The first one holds in any NCHV theory, and the second one in any LHV theory. 
	The main difference between them is that the first one is violated by any quantum state, 
	while the second is necessarily violated only by non-separable states. 
	
	\begin{figure}
		\centering
		\includegraphics[width= 0.6 \textwidth]{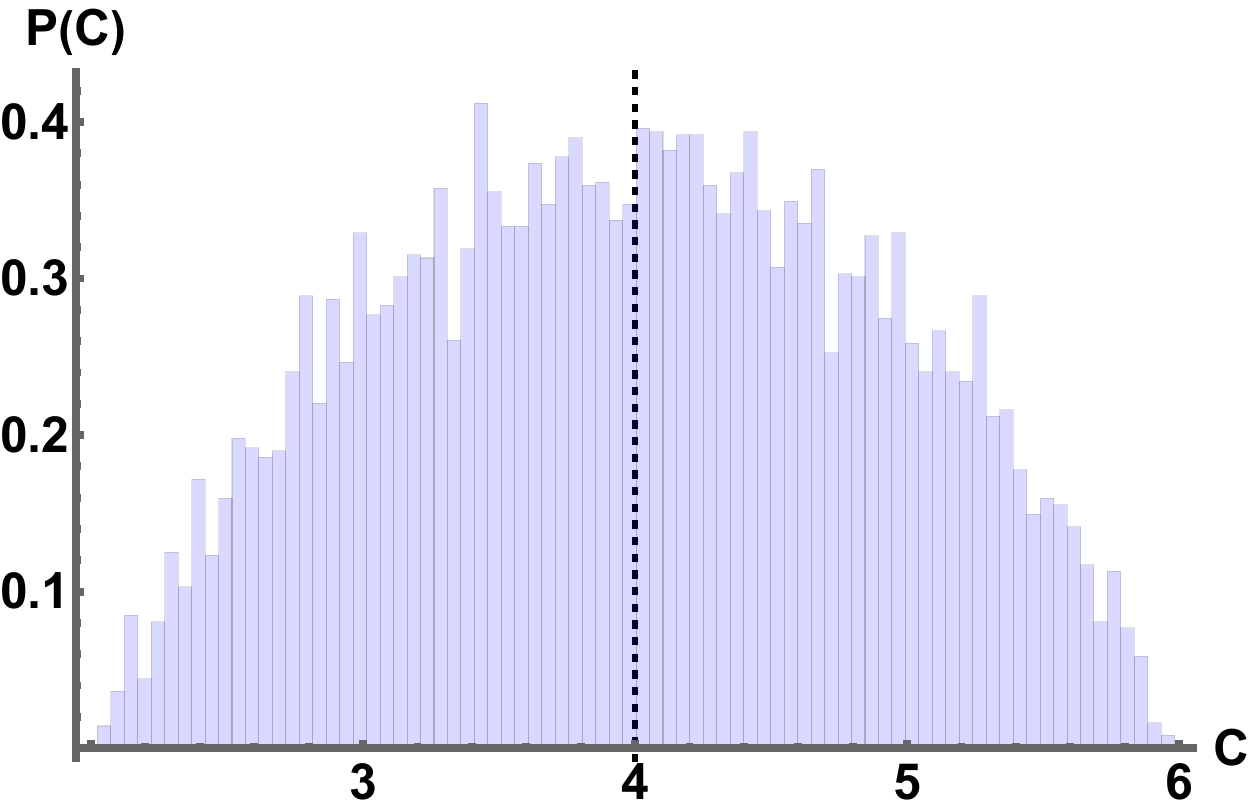}
		\caption{The probability distribution $P(C)$ of the expression in Eq.\eqref{japin} for an ensemble of 
			$M=10^6$ two-qubit random pure states. We give also the first four central moments of the distribution $\{3.998, 0.808, 0.00038, 2.132\}$.
			The black dashed line corresponds to the value separating states admitting a NCHV model $(C\leq 4)$ and those violating the contextuality inequality considered $(C>4)$. 
			Considering we are dealing with a finite number of states it is worth reporting also the median of the distribution, i.e. $m_{P(C)}= 4.007.$}
		\label{fig:1}
	\end{figure}
	
	In order to infer the nonlocal properties from the contextual one we need to identify two appropriate inequalities. 
	The author of \cite{cabello08} shows how a particular variation of the observables considered 
	by Peres and Mermin to prove the Bell-Kochen-Specker theorem can also serve as a state-dependent test of Bell nonlocality for two-qubit systems. 
	In particular, when the term $O_{33}$ is replaced with the four-dimensional identity operator the lhs of the inequality can be reformulated as: 
	\begin{eqnarray}
		\label{japin}
		C &=& \langle O_{11} O_{12} O_{13} \rangle +\langle O_{21} O_{22} O_{23} \rangle +\langle O_{31} O_{32}  \rangle \nonumber \\
		&+&    \langle O_{11} O_{21} O_{31} \rangle +\langle O_{12} O_{22} O_{32} \rangle  - \av{ O_{13} O_{23}}.
	\end{eqnarray}
	
	If we generate all the $2^{2^3}$ possible classical values of $C,$ we obtain the bound satisfied in a NCHV theory, namely $C\leq 4$. 
	To test quantum states, we use the observables as in Eq.\eqref{pmsquare}. 
	This inequality is maximally violated by the product state $\ket{\circlearrowleft}\ket{\circlearrowleft},$ where $\sigma_y\ket{\circlearrowleft} = +1 \ket{\circlearrowleft},$
	for which it reaches the value $C=6.$
	This is related to the fact that any quantum state would violate the inequality in \eqref{pmsquare} 
	and returns the same value for every quantum state due to the set of measurements considered. 
	In contrast, the noncontextual bound that follows from the expression we have introduced in Eq.\eqref{japin} allows us to perform a state-dependent test of noncontextuality. 
	\begin{figure}[H] \centering
		\includegraphics[width= 0.6 \textwidth]{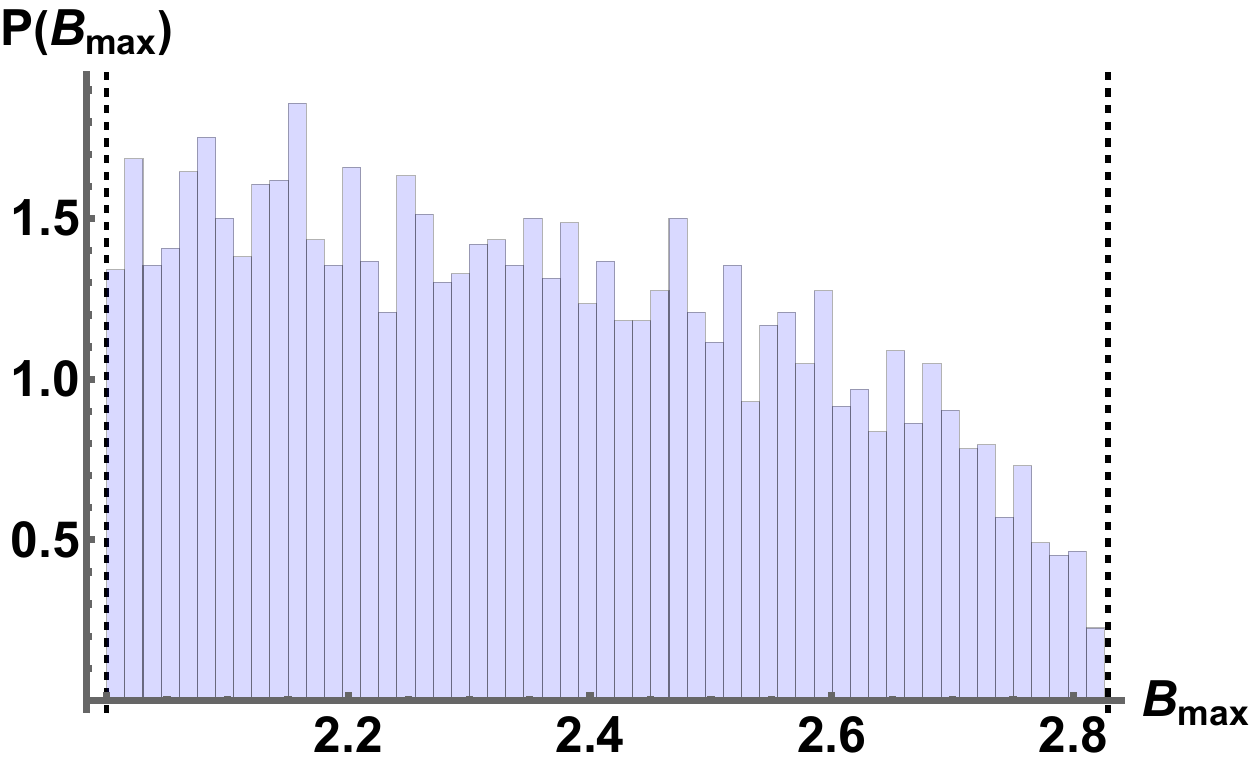}
		\caption{We report the histogram of the probability distribution $P(B_{max})$ of the expression in Equation \ref{eq:Bval} for the ensemble of 
			$M_c=50212$ two-qubit random pure states satisfying the constraint on the contextual behavior $C > 4.$ 
			The dashed black line on the left indicates the threshold value separating states admitting a local model $(B\leq 2)$ and those violating local realism $(B>2)$, 
			while the dashed black line on the right corresponds to the Tsirelson bound $(B=2\sqrt{2})$. }
		\label{fig:2}
	\end{figure}
	
	The second inequality has been already introduced in \eqref{eq:Bval}. In the case of two-qubit systems one can 
	introduce as dichotomic local observables the measurements along an arbitrary direction in the Bloch sphere of one of the two observers.  
	Therefore, the quantum mechanical observables are parametrized in the following way $W_j = \mathbf{w}_j \cdot \bm\sigma$, in which $X=\{A, B\}$ and $j=\{1,2\},$ and explicitly we have:  
	\begin{equation}
		\hat{\textbf{w}_j}\cdot \bm\sigma = \left(\begin{array}{cc}
			\cos \theta_{w_j} & e^{- i \phi_{w_j}} \sin \theta_{w_j} \\
			e^{i \phi_{w_j}} \sin \theta_{w_j} & \cos \theta_{w_j}\\
		\end{array}
		\right), 
	\end{equation}
	where the angles $\theta_{w_j}, \phi_{w_j}$ define the direction of the unit vector $\mathbf{w}_j$, 
	namely $\mathbf{w}_j= (\sin \theta_{w_j} \cos \phi_{w_j},\sin \theta_{w_j}\sin \phi_{w_j},\cos \theta_{w_j}).$
	
	Our first aim is to study the typical behavior of the value of $C$ defined in Eq.\eqref{japin}. 
	For this reason, we have generated a random ensemble of $M=10^6$ pure two-qubit states \cite{radtke2008simulation}. 
	They are obtained by choosing the columns of a random unitary matrix generated according to the Haar measure of $SU(4)$ \cite{zyczkowski1994random,cavalcanti}. 
	
	\begin{figure}[H] \centering
		\includegraphics[width= 0.6 \textwidth]{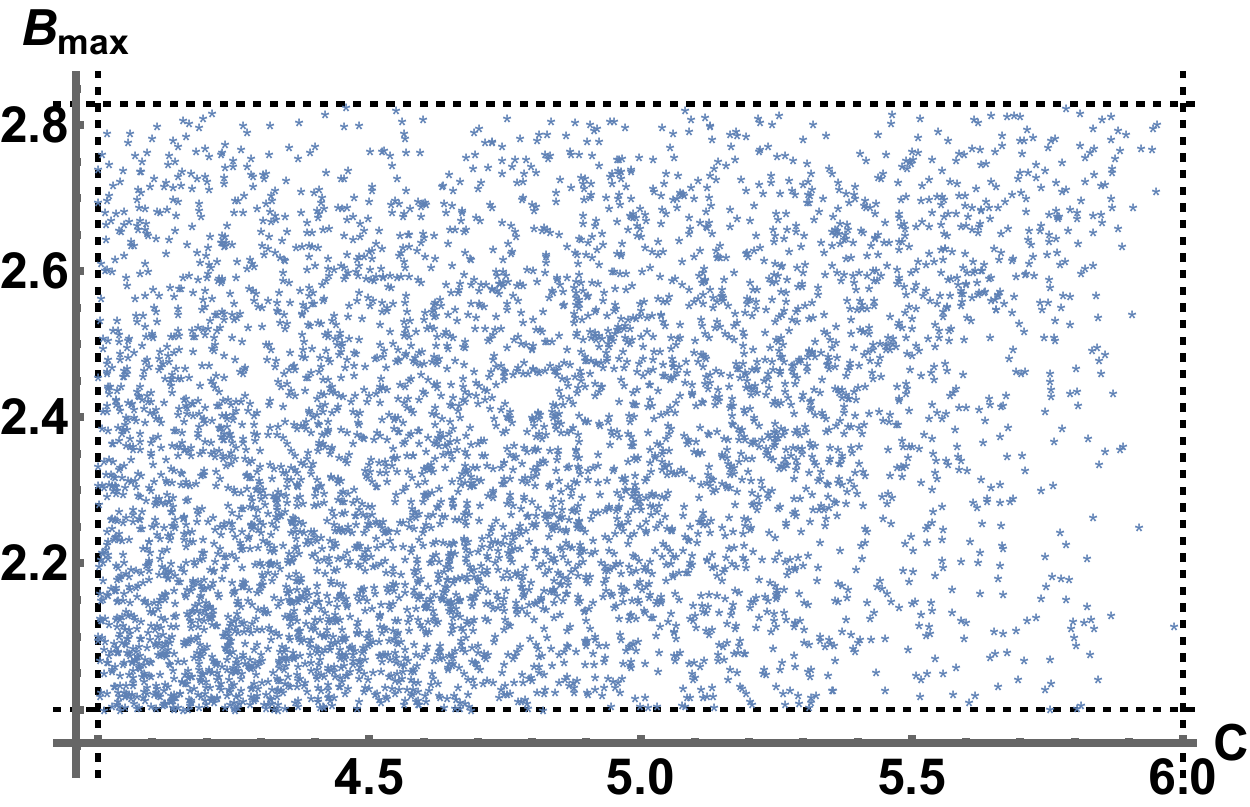}
		\caption{We show the numerical values obtained for the $M_c$ used to plot the histogram in Figure \ref{fig:2} 
			and compare for each of the quantum contextuality with the corresponding value of the maximal violation of the CHSH inequality. 
			The lower and upper horizontal black dashed lines refer to the classical and Tsirelson bound, respectively. 
			While the left and right vertical black dashed lines represent the minimum and maximal quantum contextualize,
			that we were numerically able to achieve with the functional in Equation \eqref{japin}.}
		\label{fig:3}
	\end{figure}
	
	In Figure \ref{fig:1} we report the histogram for the probability distribution $P(C)$ evaluated numerically using the ensemble of random states. 
	It is evident from our empirical evidence that the probability to have a two-qubit violating the value of 4, for which a NCHV description is possible, 
	is $P(C \geq 4) \simeq  1/2.$
	
	Subsequently, we have selected a subensamble $M_c$ of the original ensemble $M$, which is composed of the states violating the noncontextual bound. 
	We have computed the optimal violation of the expression in Equation \eqref{eq:Bval}.
	Namely, we have performed an optimization over the eight angles that defines the four directions necessary to perform the Bell test, 
	in the form of a violation of the CHSH inequality, the optimization problem can be formulated as follow: 
	and computed the maximum value: 
	\begin{equation}
		\label{eq:Bopt}
		B_{max}= \max_{\theta_{A_1}, \phi_{A_1},  \theta_{A_2},\phi_{A_2}, \theta_{B_1}, \phi_{B_1}, \theta_{B_2}, \phi_{B_2}} \vert \bra{\psi} \mathcal{B} \ket{\psi} \vert.
	\end{equation}
	
	For the optimization, we employed the RandomSearch algorithm present in the Wolfram
	Mathematica software. The choice compared with the other optimization algorithms offered 
	by the same software provides the most reliable results, and in a shorter time, see also \cite{SP_3}. The results are shown in Figure \ref{fig:2}.
	The histogram shows that all the states that are quantum contextual according to the definition given in Eq.\eqref{japin} 
	do not also admit a description in terms of local realism. Moreover, considering that we have only considered pure states, 
	the violation of the CHSH is also a signature of the nonseparability of the states and therefore, that it is entangled \cite{gisin1991bell}.
	It is worth noting now that the contextuality test has been conducted with measurements along fixed axes, 
	and does not require any numerical optimization as in the case of the CHSH test. 
	Moreover, to assess the entanglement content of a state from the experimental point of view 
	(e.g. by performing any methods based on the singular value decomposition), 
	requires a full tomography of the state, that in the case of qubit requires a quorum of observables higher \cite{fid3} than 
	those required to evaluate $C.$ 
	
	One question remains to be answered. Is it possible to infer from the value of the functional $C$ 
	how strongly a state violating the noncontextual classical will violate also the CHSH inequality? 
	In other words, we are asking if the magnitude of the violation of the classical bound of $C$ 
	for a given state $\ket{\psi}$ can give any information on how much that same state would violate the local realism. 
	The answer emerging from our numerical analysis is negative and it is reported in Figure \ref{fig:3}. 
	There, we have plotted the value of $B_{max}$ as a function of the corresponding value of $C$
	for all the $M_c$  states. This analysis confirms that contextuality and Bell nonlocality are independent concepts unless further studies towards this direction would unveil the existence of an underpinned mechanism that activates an emergent fundamental connection between them.
	
	\section{Conclusions and applications}
	The forthcoming success of quantum technologies like cryptography, computing, and sensing is inextricably linked to advancements at the most fundamental level. 
	Therefore expanding the theoretical and foundational basis for the design of future quantum technologies is nowadays of paramount importance.  
	In particular, we studied the relationship between quantum contextuality and Bell-nonclassicality for an ensemble of two-qubit pure states. 
	We focused on pure states because, in the near future we expect to move from noisy quantum devices to fault-tolerant ones. 
	Therefore, the need to devise a test for evaluating effective quantum resources is increasingly essential.
	Therefore, the need to devise a test for evaluating effective quantum resources and the connection among them \cite{Diker_2022} is essential. 
	Drawing inspiration from one of the most straightforward proofs of quantum contextuality - the Peres-Mermin square - we managed to establish a link between a simple inequality targeted at demonstrating quantum contextuality and the impossibility of depicting the state in a local realistic scenario.
	
	Our analysis based on typical pure quantum states seems to suggest that the violation of the noncontextual bound of Eq. \eqref{japin} suggest to be 
	a sufficient condition for the subsequent violation of the CHSH inequality. Anyway, we are aware that this cannot be true 
	due to the fact that a separable state maximally violates the aforementioned bound. 
	In our opinion, a deeper analysis, outside the scope of the present paper, is to investigate the relationship between the local polytope \cite{gigena2022a} 
	defined in the CHSH scenario and the noncontextual one.  
	We acknowledge that experiments testing Bell inequalities on different degrees of freedom of the same particle, 
	have to be intended as a way to rule out NCHV theories and not the LHV models \cite{dambrioso_contex}.
	In contrast, our aim is complementary to those already established and we numerically addressed how the detection of 
	the violation of a suitably chosen Bell-like inequality for contextuality would give a hint on the amount of ``truly'' Bell-nonclassicality of the state.

	\section*{Acknowledgments}
	GS is supported by QuantERA/2/2020, an ERA-Net co-fund in Quantum Technologies, under the eDICT project. 
	AM is supported by  Foundation for Polish Science (FNP), IRAP project ICTQT, contract no. 2018/MAB/5, co-financed by EU  Smart Growth Operational Programme.


	
	
	
	
	
	\bibliography{bib_IJTP.bib}


\begin{thebibliography}{42}
\ifx \bisbn   \undefined \def \bisbn  #1{ISBN #1}\fi
\ifx \binits  \undefined \def \binits#1{#1}\fi
\ifx \bauthor  \undefined \def \bauthor#1{#1}\fi
\ifx \batitle  \undefined \def \batitle#1{#1}\fi
\ifx \bjtitle  \undefined \def \bjtitle#1{#1}\fi
\ifx \bvolume  \undefined \def \bvolume#1{\textbf{#1}}\fi
\ifx \byear  \undefined \def \byear#1{#1}\fi
\ifx \bissue  \undefined \def \bissue#1{#1}\fi
\ifx \bfpage  \undefined \def \bfpage#1{#1}\fi
\ifx \blpage  \undefined \def \blpage #1{#1}\fi
\ifx \burl  \undefined \def \burl#1{\textsf{#1}}\fi
\ifx \doiurl  \undefined \def \doiurl#1{\url{https://doi.org/#1}}\fi
\ifx \betal  \undefined \def \betal{\textit{et al.}}\fi
\ifx \binstitute  \undefined \def \binstitute#1{#1}\fi
\ifx \binstitutionaled  \undefined \def \binstitutionaled#1{#1}\fi
\ifx \bctitle  \undefined \def \bctitle#1{#1}\fi
\ifx \beditor  \undefined \def \beditor#1{#1}\fi
\ifx \bpublisher  \undefined \def \bpublisher#1{#1}\fi
\ifx \bbtitle  \undefined \def \bbtitle#1{#1}\fi
\ifx \bedition  \undefined \def \bedition#1{#1}\fi
\ifx \bseriesno  \undefined \def \bseriesno#1{#1}\fi
\ifx \blocation  \undefined \def \blocation#1{#1}\fi
\ifx \bsertitle  \undefined \def \bsertitle#1{#1}\fi
\ifx \bsnm \undefined \def \bsnm#1{#1}\fi
\ifx \bsuffix \undefined \def \bsuffix#1{#1}\fi
\ifx \bparticle \undefined \def \bparticle#1{#1}\fi
\ifx \barticle \undefined \def \barticle#1{#1}\fi
\bibcommenthead
\ifx \bconfdate \undefined \def \bconfdate #1{#1}\fi
\ifx \botherref \undefined \def \botherref #1{#1}\fi
\ifx \url \undefined \def \url#1{\textsf{#1}}\fi
\ifx \bchapter \undefined \def \bchapter#1{#1}\fi
\ifx \bbook \undefined \def \bbook#1{#1}\fi
\ifx \bcomment \undefined \def \bcomment#1{#1}\fi
\ifx \oauthor \undefined \def \oauthor#1{#1}\fi
\ifx \citeauthoryear \undefined \def \citeauthoryear#1{#1}\fi
\ifx \endbibitem  \undefined \def \endbibitem {}\fi
\ifx \bconflocation  \undefined \def \bconflocation#1{#1}\fi
\ifx \arxivurl  \undefined \def \arxivurl#1{\textsf{#1}}\fi
\csname PreBibitemsHook\endcsname

\bibitem{RBell_C}
\begin{barticle}
\bauthor{\bsnm{BELL}, \binits{J.S.}}:
\batitle{On the problem of hidden variables in quantum mechanics}.
\bjtitle{Rev. Mod. Phys.}
\bvolume{38},
\bfpage{447}--\blpage{452}
(\byear{1966}).
\doiurl{10.1103/RevModPhys.38.447}
\end{barticle}
\endbibitem

\bibitem{KS_th}
\begin{barticle}
\bauthor{\bsnm{KOCHEN}, \binits{S.}},
\bauthor{\bsnm{SPECKER}, \binits{E.P.}}:
\batitle{The problem of hidden variables in quantum mechanics}.
\bjtitle{Journal of Mathematics and Mechanics}
\bvolume{17}(\bissue{1}),
\bfpage{59}--\blpage{87}
(\byear{1967})
\end{barticle}
\endbibitem

\bibitem{Bell_th}
\begin{barticle}
\bauthor{\bsnm{Bell}, \binits{J.S.}}:
\batitle{On the einstein podolsky rosen paradox}.
\bjtitle{Physics Physique Fizika}
\bvolume{1},
\bfpage{195}--\blpage{200}
(\byear{1964}).
\doiurl{10.1103/PhysicsPhysiqueFizika.1.195}
\end{barticle}
\endbibitem

\bibitem{CHSH}
\begin{barticle}
\bauthor{\bsnm{Clauser}, \binits{J.F.}},
\bauthor{\bsnm{Horne}, \binits{M.A.}},
\bauthor{\bsnm{Shimony}, \binits{A.}},
\bauthor{\bsnm{Holt}, \binits{R.A.}}:
\batitle{Proposed experiment to test local hidden-variable theories}.
\bjtitle{Phys. Rev. Lett.}
\bvolume{23},
\bfpage{880}--\blpage{884}
(\byear{1969}).
\doiurl{10.1103/PhysRevLett.23.880}
\end{barticle}
\endbibitem

\bibitem{CH74}
\begin{barticle}
\bauthor{\bsnm{Clauser}, \binits{J.F.}},
\bauthor{\bsnm{Horne}, \binits{M.A.}}:
\batitle{Experimental consequences of objective local theories}.
\bjtitle{Phys. Rev. D}
\bvolume{10},
\bfpage{526}--\blpage{535}
(\byear{1974}).
\doiurl{10.1103/PhysRevD.10.526}
\end{barticle}
\endbibitem

\bibitem{Zukowski2002}
\begin{barticle}
\bauthor{\bsnm{{\.{Z}}ukowski}, \binits{M.}},
\bauthor{\bsnm{Brukner}, \binits{{\v{C}}.}}:
\batitle{Bell's theorem for general $n$-qubit states}.
\bjtitle{Physical Review Letters}
\bvolume{88}(\bissue{21}),
\bfpage{210401}
(\byear{2002}).
\doiurl{10.1103/physrevlett.88.210401}
\end{barticle}
\endbibitem

\bibitem{Karczewski2022}
\begin{barticle}
\bauthor{\bsnm{Karczewski}, \binits{M.}},
\bauthor{\bsnm{Scala}, \binits{G.}},
\bauthor{\bsnm{Mandarino}, \binits{A.}},
\bauthor{\bsnm{Sainz}, \binits{A.B.}},
\bauthor{\bsnm{{\.{Z}}ukowski}, \binits{M.}}:
\batitle{Avenues to generalising bell inequalities}.
\bjtitle{Journal of Physics A: Mathematical and Theoretical}
\bvolume{55}(\bissue{38}),
\bfpage{384011}
(\byear{2022}).
\doiurl{10.1088/1751-8121/ac8a28}
\end{barticle}
\endbibitem

\bibitem{Bae2018}
\begin{barticle}
\bauthor{\bsnm{Bae}, \binits{K.}},
\bauthor{\bsnm{Son}, \binits{W.}}:
\batitle{Generalized nonlocality criteria under the correlation symmetry}.
\bjtitle{Physical Review A}
\bvolume{98}(\bissue{2}),
\bfpage{022116}
(\byear{2018}).
\doiurl{10.1103/physreva.98.022116}
\end{barticle}
\endbibitem

\bibitem{cabello08}
\begin{barticle}
\bauthor{\bsnm{Cabello}, \binits{A.}}:
\batitle{Experimentally testable state-independent quantum contextuality}.
\bjtitle{Phys. Rev. Lett.}
\bvolume{101},
\bfpage{210401}
(\byear{2008}).
\doiurl{10.1103/PhysRevLett.101.210401}
\end{barticle}
\endbibitem

\bibitem{KCBS}
\begin{barticle}
\bauthor{\bsnm{Klyachko}, \binits{A.A.}},
\bauthor{\bsnm{Can}, \binits{M.A.}},
\bauthor{\bsnm{Binicio{\u{g}}lu}, \binits{S.}},
\bauthor{\bsnm{Shumovsky}, \binits{A.S.}}:
\batitle{Simple test for hidden variables in spin-1 systems}.
\bjtitle{Phys. Rev. Lett.}
\bvolume{101},
\bfpage{020403}
(\byear{2008}).
\doiurl{10.1103/PhysRevLett.101.020403}
\end{barticle}
\endbibitem

\bibitem{DeviceInd_rev}
\begin{barticle}
\bauthor{\bsnm{Primaatmaja}, \binits{I.W.}},
\bauthor{\bsnm{Goh}, \binits{K.T.}},
\bauthor{\bsnm{Tan}, \binits{E.Y.-Z.}},
\bauthor{\bsnm{Khoo}, \binits{J.T.-F.}},
\bauthor{\bsnm{Ghorai}, \binits{S.}},
\bauthor{\bsnm{Lim}, \binits{C.C.-W.}}:
\batitle{Security of device-independent quantum key distribution protocols: a
  review}.
\bjtitle{Quantum}
\bvolume{7},
\bfpage{932}
(\byear{2023})
\end{barticle}
\endbibitem

\bibitem{budroni2022kochen}
\begin{barticle}
\bauthor{\bsnm{Budroni}, \binits{C.}},
\bauthor{\bsnm{Cabello}, \binits{A.}},
\bauthor{\bsnm{G\"uhne}, \binits{O.}},
\bauthor{\bsnm{Kleinmann}, \binits{M.}},
\bauthor{\bsnm{Larsson}, \binits{J.-A.k.}}:
\batitle{Kochen-specker contextuality}.
\bjtitle{Rev. Mod. Phys.}
\bvolume{94},
\bfpage{045007}
(\byear{2022}).
\doiurl{10.1103/RevModPhys.94.045007}
\end{barticle}
\endbibitem

\bibitem{Ekert1991}
\begin{barticle}
\bauthor{\bsnm{Ekert}, \binits{A.K.}}:
\batitle{Quantum cryptography based on {B}ell's theorem}.
\bjtitle{Phys. Rev. Lett.}
\bvolume{67},
\bfpage{661}--\blpage{663}
(\byear{1991}).
\doiurl{10.1103/PhysRevLett.67.661}
\end{barticle}
\endbibitem

\bibitem{hyperbit}
\begin{botherref}
\oauthor{\bsnm{Scala}, \binits{G.}},
\oauthor{\bsnm{Ghoreishi}, \binits{S.A.}},
\oauthor{\bsnm{Paw{\l}owski}, \binits{M.}}:
Revisiting hyperbit limitations unveils quantum communication advantages.
arXiv preprint arXiv:2308.16114
(2023)
\end{botherref}
\endbibitem

\bibitem{Pepe_2022}
\begin{botherref}
\oauthor{\bsnm{Pepe}, \binits{F.V.}},
\oauthor{\bsnm{Scala}, \binits{G.}},
\oauthor{\bsnm{Chilleri}, \binits{G.}},
\oauthor{\bsnm{Triggiani}, \binits{D.}},
\oauthor{\bsnm{Kim}, \binits{Y.-H.}},
\oauthor{\bsnm{Tamma}, \binits{V.}}:
Distance sensitivity of thermal light second-order interference beyond spatial
  coherence.
The European Physical Journal Plus
\textbf{137}(6)
(2022).
\doiurl{10.1140/epjp/s13360-022-02857-7}
\end{botherref}
\endbibitem

\bibitem{1stPaper}
\begin{barticle}
\bauthor{\bsnm{Das}, \binits{T.}},
\bauthor{\bsnm{Karczewski}, \binits{M.}},
\bauthor{\bsnm{Mandarino}, \binits{A.}},
\bauthor{\bsnm{Markiewicz}, \binits{M.}},
\bauthor{\bsnm{Woloncewicz}, \binits{B.}},
\bauthor{\bsnm{{\.{Z}}ukowski}, \binits{M.}}:
\batitle{Wave{\textendash}particle complementarity: detecting violation of
  local realism with photon-number resolving weak-field homodyne measurements}.
\bjtitle{New Journal of Physics}
\bvolume{24}(\bissue{3}),
\bfpage{033017}
(\byear{2022}).
\doiurl{10.1088/1367-2630/ac54c8}
\end{barticle}
\endbibitem

\bibitem{1stPLA}
\begin{barticle}
\bauthor{\bsnm{Das}, \binits{T.}},
\bauthor{\bsnm{Karczewski}, \binits{M.}},
\bauthor{\bsnm{Mandarino}, \binits{A.}},
\bauthor{\bsnm{Markiewicz}, \binits{M.}},
\bauthor{\bsnm{Woloncewicz}, \binits{B.}},
\bauthor{\bsnm{Żukowski}, \binits{M.}}:
\batitle{Remarks about bell-nonclassicality of a single photon}.
\bjtitle{Physics Letters A}
\bvolume{435},
\bfpage{128031}
(\byear{2022}).
\doiurl{10.1016/j.physleta.2022.128031}
\end{barticle}
\endbibitem

\bibitem{2ndPaper}
\begin{barticle}
\bauthor{\bsnm{Das}, \binits{T.}},
\bauthor{\bsnm{Karczewski}, \binits{M.}},
\bauthor{\bsnm{Mandarino}, \binits{A.}},
\bauthor{\bsnm{Markiewicz}, \binits{M.}},
\bauthor{\bsnm{Woloncewicz}, \binits{B.}},
\bauthor{\bsnm{{\.{Z}}ukowski}, \binits{M.}}:
\batitle{Can single photon excitation of two spatially separated modes lead to
  a violation of bell inequality via weak-field homodyne measurements?}
\bjtitle{New Journal of Physics}
\bvolume{23}(\bissue{7}),
\bfpage{073042}
(\byear{2021}).
\doiurl{10.1088/1367-2630/ac0ffe}
\end{barticle}
\endbibitem

\bibitem{CommentDun}
\begin{barticle}
\bauthor{\bsnm{Das}, \binits{T.}},
\bauthor{\bsnm{Karczewski}, \binits{M.}},
\bauthor{\bsnm{Mandarino}, \binits{A.}},
\bauthor{\bsnm{Markiewicz}, \binits{M.}},
\bauthor{\bsnm{Woloncewicz}, \binits{B.}},
\bauthor{\bsnm{{\.{Z}}ukowski}, \binits{M.}}:
\batitle{Comment on `single particle nonlocality with completely independent
  reference states'}.
\bjtitle{New Journal of Physics}
\bvolume{24}(\bissue{3}),
\bfpage{038001}
(\byear{2022}).
\doiurl{10.1088/1367-2630/ac55b1}
\end{barticle}
\endbibitem

\bibitem{Catani2022a}
\begin{botherref}
\oauthor{\bsnm{Catani}, \binits{L.}},
\oauthor{\bsnm{Leifer}, \binits{M.}},
\oauthor{\bsnm{Scala}, \binits{G.}},
\oauthor{\bsnm{Schmid}, \binits{D.}},
\oauthor{\bsnm{Spekkens}, \binits{R.W.}}:
What aspects of the phenomenology of interference witness nonclassicality?
(2022)
{\href{https://arxiv.org/abs/2211.09850}{{arXiv:2211.09850}}}
{[quant-ph]}
\end{botherref}
\endbibitem

\bibitem{Catani2022b}
\begin{botherref}
\oauthor{\bsnm{Catani}, \binits{L.}},
\oauthor{\bsnm{Leifer}, \binits{M.}},
\oauthor{\bsnm{Scala}, \binits{G.}},
\oauthor{\bsnm{Schmid}, \binits{D.}},
\oauthor{\bsnm{Spekkens}, \binits{R.W.}}:
What is nonclassical about uncertainty relations?
Physical Review Letters
\textbf{129}(24)
(2022).
\doiurl{10.1103/physrevlett.129.240401}
\end{botherref}
\endbibitem

\bibitem{peres1990incompatible}
\begin{barticle}
\bauthor{\bsnm{Peres}, \binits{A.}}:
\batitle{Incompatible results of quantum measurements}.
\bjtitle{Physics Letters A}
\bvolume{151}(\bissue{3-4}),
\bfpage{107}--\blpage{108}
(\byear{1990})
\end{barticle}
\endbibitem

\bibitem{mermin90}
\begin{barticle}
\bauthor{\bsnm{Mermin}, \binits{N.D.}}:
\batitle{Simple unified form for the major no-hidden-variables theorems}.
\bjtitle{Phys. Rev. Lett.}
\bvolume{65},
\bfpage{3373}--\blpage{3376}
(\byear{1990}).
\doiurl{10.1103/PhysRevLett.65.3373}
\end{barticle}
\endbibitem

\bibitem{peres1991two}
\begin{barticle}
\bauthor{\bsnm{Peres}, \binits{A.}}:
\batitle{Two simple proofs of the kochen-specker theorem}.
\bjtitle{Journal of Physics A: Mathematical and General}
\bvolume{24}(\bissue{4}),
\bfpage{175}
(\byear{1991})
\end{barticle}
\endbibitem

\bibitem{peres1992recursive}
\begin{barticle}
\bauthor{\bsnm{Peres}, \binits{A.}}:
\batitle{Recursive definition for elements of reality}.
\bjtitle{Foundations of physics}
\bvolume{22},
\bfpage{357}--\blpage{361}
(\byear{1992})
\end{barticle}
\endbibitem

\bibitem{Exp_context1}
\begin{barticle}
\bauthor{\bsnm{Simon}, \binits{C.}},
\bauthor{\bparticle{\ifmmode~\dot{Z}\else} \bsnm{\.{Z}\fi{}ukowski},
  \binits{M.}},
\bauthor{\bsnm{Weinfurter}, \binits{H.}},
\bauthor{\bsnm{Zeilinger}, \binits{A.}}:
\batitle{Feasible ``kochen-specker'' experiment with single particles}.
\bjtitle{Phys. Rev. Lett.}
\bvolume{85},
\bfpage{1783}--\blpage{1786}
(\byear{2000}).
\doiurl{10.1103/PhysRevLett.85.1783}
\end{barticle}
\endbibitem

\bibitem{Exp_context2}
\begin{barticle}
\bauthor{\bsnm{Amselem}, \binits{E.}},
\bauthor{\bsnm{R\aa{}dmark}, \binits{M.}},
\bauthor{\bsnm{Bourennane}, \binits{M.}},
\bauthor{\bsnm{Cabello}, \binits{A.}}:
\batitle{State-independent quantum contextuality with single photons}.
\bjtitle{Phys. Rev. Lett.}
\bvolume{103},
\bfpage{160405}
(\byear{2009}).
\doiurl{10.1103/PhysRevLett.103.160405}
\end{barticle}
\endbibitem

\bibitem{Non_contexAM}
\begin{barticle}
\bauthor{\bsnm{Schlichtholz}, \binits{K.}},
\bauthor{\bsnm{Mandarino}, \binits{A.}},
\bauthor{\bsnm{{\.{Z}}ukowski}, \binits{M.}}:
\batitle{Bosonic fields in states with undefined particle numbers possess
  detectable non-contextuality features, plus more}.
\bjtitle{New Journal of Physics}
\bvolume{24}(\bissue{10}),
\bfpage{103003}
(\byear{2022}).
\doiurl{10.1088/1367-2630/ac0ffe}
\end{barticle}
\endbibitem

\bibitem{cabello98}
\begin{barticle}
\bauthor{\bsnm{Cabello}, \binits{A.}},
\bauthor{\bsnm{Garc\'{\i}a-Alcaine}, \binits{G.}}:
\batitle{Proposed experimental tests of the bell-kochen-specker theorem}.
\bjtitle{Phys. Rev. Lett.}
\bvolume{80},
\bfpage{1797}--\blpage{1799}
(\byear{1998}).
\doiurl{10.1103/PhysRevLett.80.1797}
\end{barticle}
\endbibitem

\bibitem{cabello_exp08}
\begin{barticle}
\bauthor{\bsnm{Cabello}, \binits{A.}},
\bauthor{\bsnm{Filipp}, \binits{S.}},
\bauthor{\bsnm{Rauch}, \binits{H.}},
\bauthor{\bsnm{Hasegawa}, \binits{Y.}}:
\batitle{Proposed experiment for testing quantum contextuality with neutrons}.
\bjtitle{Phys. Rev. Lett.}
\bvolume{100},
\bfpage{130404}
(\byear{2008}).
\doiurl{10.1103/PhysRevLett.100.130404}
\end{barticle}
\endbibitem

\bibitem{cabello_exp09}
\begin{barticle}
\bauthor{\bsnm{Bartosik}, \binits{H.}},
\bauthor{\bsnm{Klepp}, \binits{J.}},
\bauthor{\bsnm{Schmitzer}, \binits{C.}},
\bauthor{\bsnm{Sponar}, \binits{S.}},
\bauthor{\bsnm{Cabello}, \binits{A.}},
\bauthor{\bsnm{Rauch}, \binits{H.}},
\bauthor{\bsnm{Hasegawa}, \binits{Y.}}:
\batitle{Experimental test of quantum contextuality in neutron interferometry}.
\bjtitle{Phys. Rev. Lett.}
\bvolume{103},
\bfpage{040403}
(\byear{2009}).
\doiurl{10.1103/PhysRevLett.103.040403}
\end{barticle}
\endbibitem

\bibitem{gigena2022a}
\begin{botherref}
\oauthor{\bsnm{Gigena}, \binits{N.}},
\oauthor{\bsnm{Scala}, \binits{G.}},
\oauthor{\bsnm{Mandarino}, \binits{A.}}:
Revisited aspects of the local set in chsh bell scenario.
International Journal of Quantum Information.
\doiurl{10.1142/S0219749923400051}
\end{botherref}
\endbibitem

\bibitem{entropy_FID_CHSH}
\begin{barticle}
\bauthor{\bsnm{Mandarino}, \binits{A.}},
\bauthor{\bsnm{Scala}, \binits{G.}}:
\batitle{On the fidelity robustness of chsh--bell inequality via filtered
  random states}.
\bjtitle{Entropy}
\bvolume{25}(\bissue{1}),
\bfpage{94}
(\byear{2023}).
\doiurl{10.3390/e25010094}
\end{barticle}
\endbibitem

\bibitem{Cirelson1980}
\begin{barticle}
\bauthor{\bsnm{Cirel{\textquotesingle}son}, \binits{B.S.}}:
\batitle{Quantum generalizations of bell{\textquotesingle}s inequality}.
\bjtitle{Letters in Mathematical Physics}
\bvolume{4}(\bissue{2}),
\bfpage{93}--\blpage{100}
(\byear{1980}).
\doiurl{10.1007/bf00417500}
\end{barticle}
\endbibitem

\bibitem{radtke2008simulation}
\begin{barticle}
\bauthor{\bsnm{Radtke}, \binits{T.}},
\bauthor{\bsnm{Fritzsche}, \binits{S.}}:
\batitle{Simulation of n-qubit quantum systems. iv. parametrizations of quantum
  states, matrices and probability distributions}.
\bjtitle{Computer Physics Communications}
\bvolume{179}(\bissue{9}),
\bfpage{647}--\blpage{664}
(\byear{2008})
\end{barticle}
\endbibitem

\bibitem{zyczkowski1994random}
\begin{barticle}
\bauthor{\bsnm{Zyczkowski}, \binits{K.}},
\bauthor{\bsnm{Kus}, \binits{M.}}:
\batitle{Random unitary matrices}.
\bjtitle{Journal of Physics A: Mathematical and General}
\bvolume{27}(\bissue{12}),
\bfpage{4235}
(\byear{1994})
\end{barticle}
\endbibitem

\bibitem{cavalcanti}
\begin{botherref}
\oauthor{\bsnm{Cavalcanti}, \binits{P.J.}},
\oauthor{\bsnm{Scala}, \binits{G.}},
\oauthor{\bsnm{Mandarino}, \binits{A.}},
\oauthor{\bsnm{Lupo}, \binits{C.}}:
Information theoretical perspective on the method of entanglement witnesses.
arXiv preprint arXiv:2308.07744
(2023)
\end{botherref}
\endbibitem

\bibitem{SP_3}
\begin{barticle}
\bauthor{\bsnm{Das}, \binits{T.}},
\bauthor{\bsnm{Karczewski}, \binits{M.}},
\bauthor{\bsnm{Mandarino}, \binits{A.}},
\bauthor{\bsnm{Markiewicz}, \binits{M.}},
\bauthor{\bparticle{\ifmmode~\dot{Z}\else} \bsnm{\.{Z}\fi{}ukowski},
  \binits{M.}}:
\batitle{Optimal interferometry for bell nonclassicality induced by a
  vacuum--one-photon qubit}.
\bjtitle{Phys. Rev. Applied}
\bvolume{18},
\bfpage{034074}
(\byear{2022}).
\doiurl{10.1103/PhysRevApplied.18.034074}
\end{barticle}
\endbibitem

\bibitem{gisin1991bell}
\begin{barticle}
\bauthor{\bsnm{Gisin}, \binits{N.}}:
\batitle{Bell's inequality holds for all non-product states}.
\bjtitle{Physics Letters A}
\bvolume{154}(\bissue{5-6}),
\bfpage{201}--\blpage{202}
(\byear{1991})
\end{barticle}
\endbibitem

\bibitem{fid3}
\begin{barticle}
\bauthor{\bsnm{Mandarino}, \binits{A.}},
\bauthor{\bsnm{Bina}, \binits{M.}},
\bauthor{\bsnm{Porto}, \binits{C.}},
\bauthor{\bsnm{Cialdi}, \binits{S.}},
\bauthor{\bsnm{Olivares}, \binits{S.}},
\bauthor{\bsnm{Paris}, \binits{M.G.A.}}:
\batitle{Assessing the significance of fidelity as a figure of merit in quantum
  state reconstruction of discrete and continuous-variable systems}.
\bjtitle{Phys. Rev. A}
\bvolume{93},
\bfpage{062118}
(\byear{2016}).
\doiurl{10.1103/PhysRevA.93.062118}
\end{barticle}
\endbibitem

\bibitem{Diker_2022}
\begin{botherref}
\oauthor{\bsnm{Diker}, \binits{F.}},
\oauthor{\bsnm{Gedik}, \binits{Z.}}:
The degree of quantum contextuality in terms of concurrence for the {KCBS}
  scenario.
International Journal of Theoretical Physics
\textbf{61}(11)
(2022).
\doiurl{10.1007/s10773-022-05245-0}
\end{botherref}
\endbibitem

\bibitem{dambrioso_contex}
\begin{barticle}
\bauthor{\bsnm{D'Ambrosio}, \binits{V.}},
\bauthor{\bsnm{Herbauts}, \binits{I.}},
\bauthor{\bsnm{Amselem}, \binits{E.}},
\bauthor{\bsnm{Nagali}, \binits{E.}},
\bauthor{\bsnm{Bourennane}, \binits{M.}},
\bauthor{\bsnm{Sciarrino}, \binits{F.}},
\bauthor{\bsnm{Cabello}, \binits{A.}}:
\batitle{Experimental implementation of a kochen-specker set of quantum tests}.
\bjtitle{Phys. Rev. X}
\bvolume{3},
\bfpage{011012}
(\byear{2013}).
\doiurl{10.1103/PhysRevX.3.011012}
\end{barticle}
\endbibitem

\end{thebibliography}
	
\end{document}